\documentstyle[aps,12pt,psfig,epsfig,floats,citesort]{revtex}

\newcommand{\ageq}{\mbox{\
\raisebox{-.9ex}{$\stackrel{\textstyle >}{\sim}$}\ }}

\def\eps{{\epsilon}}

\def\begineq{\begin{equation}}
\def\endeq{\end{equation}}
\def\be{\begin{equation}}
\def\ee{\end{equation}}

\begin{document}
\bibliographystyle{prsty}

\title{Thermal convection for large Prandtl numbers}
\author{Siegfried Grossmann$^1$ and Detlef Lohse$^2$}
\address{
$^1$ Department of Physics, University of Marburg, Renthof 6, 
D-35032 Marburg, Germany\\
$^2$Department of Applied Physics  and J.\ M.\ Burgers Centre for
Fluid Dynamics, University of Twente, 7500 AE Enschede, Netherlands\\
}

\date{\today}

\maketitle
\begin{abstract}
The Rayleigh-Benard theory by Grossmann and Lohse
[J.\ Fluid Mech.\ 407, 27 (2000)] is extended towards very large
Prandtl numbers $Pr$. The Nusselt number 
$Nu$ is found here to be independent of $Pr$. However, 
for fixed Rayleigh numbers $Ra> 10^{10}$ a maximum around
$Pr\approx 2$ in the $Nu(Pr)$-dependence is predicted which is
absent for lower $Ra$. 
We moreover offer the full functional dependences of  
$Nu(Ra,Pr)$ and $Re(Ra,Pr)$ within this extended theory, 
rather than only giving
the limiting power laws as done in ref.\ \cite{gro00}. This enables us 
to more realistically describe the {\it transitions}  
between the various scaling regimes, including their widths. 
\end{abstract}

\vspace{0.8cm}

In thermal convection, the  control parameters are the 
Rayleigh number $Ra$ and the Prandtl number $Pr$. The system responds
with the Nusselt number $Nu$ (the dimensionless heat flux) and
the Reynolds number $Re$ (the dimensionless large scale velocity). 
The key question is to understand the dependences
 $Nu(Ra,Pr)$ and $Re(Ra,Pr)$.
In experiments, traditionally the Prandtl number 
was more or less kept fixed \cite{cas89,sig94,cio97}. 
However, the recent experiments in the vicinity of the
critical point of helium gas \cite{cha97,nie00} and of
SF$_6$ \cite{ash99} or with various alcohols \cite{ahl00} allow to vary
both $Ra$ and $Pr$ and thus to explore a larger domain of the $Ra-Pr$ parameter
space of Rayleigh-Benard (RB) convection, in particular that for 
$Pr\gg 1$. While the experiments of Steinberg's group \cite{ash99}
suggest a decreasing Nusselt number with increasing $Pr$, namely
$Nu = 0.22 Ra^{0.3\pm 0.03 } Pr^{-0.2 \pm 0.04}$ in
$10^9 \le Ra \le 10^{14}$ and $1\le Pr \le 93$, the experiments of the
Ahlers group suggest a saturation of $Nu$ with increasing $Pr$ for fixed $Ra$,
at least up to $Ra=10^{10}$ \cite{ahlers_sb00}. 
The same saturation (at fixed $Ra=6\cdot 10^5$) is found in the numerical
simulations by Verzicco and Camussi \cite{ver99} and Herring and Kerr
\cite{her00}.

The large $Pr$ regime
 of the latest experiments has not been covered by the 
recent theory on thermal convection by Grossmann and Lohse (GL, \cite{gro00}),
which otherwise does pretty well in accounting for various measurements.
In particular, it explains the low $Pr$ measurements of Cioni et al.\ 
\cite{cio97} ($Pr=0.025$),
the low $Pr$ numerics  which reveal
$Nu\sim Pr^{0.14}$ for fixed $Ra$ \cite{ver99,her00},
 and the above mentioned experiments by 
Niemela et al.\ \cite{nie00} and Xu et al.\ \cite{ahl00}. 

In the present paper we extend the GL theory in a natural way to the
regime of very large $Pr$, on which no statement has been
made in the original paper  \cite{gro00}. We find $Nu$ to be 
independent of $Pr$ in that regime. 
We 
in addition 
present 
the complete functional dependences $Nu(Ra,Pr)$ and $Re(Ra,Pr)$ 
within the GL theory, rather than only giving the limiting power laws 
and superpositions of those as was done in  \cite{gro00}. This enables 
us to more realistically describe the {\it transitions} between the various 
scaling regimes found already in  \cite{gro00}.

{\it Approach:} 
To make this paper selfcontained we very briefly recapitulate the
key idea of the GL theory, which is to decompose  in the
volume averages of the energy dissipation rate $\eps_u$ and the thermal 
dissipation rate $\eps_\theta$ into their boundary layer (BL) and 
bulk contributions,
\begin{eqnarray}
\eps_u &=& \eps_{u,BL} + \eps_{u,bulk}, \label{eq1}\\
\eps_\theta &=& \eps_{\theta ,BL} + \eps_{\theta , bulk}. \label{eq2}
\end{eqnarray}
For the left hand sides the exact relations 
$
\eps_u = {\nu^3\over L^4 } (Nu-1) Ra Pr^{-2}
$
and
$
\eps_\theta = \kappa {\Delta^2 \over L^2} Nu
$ are used, where $\nu$ is the kinematic viscosity, $\kappa$ the thermal 
diffusivity, $L$ the height of the cell, and $\Delta$ the temperature 
difference between the bottom and the top plates. Next, the local dissipation 
rates in the BL and in the bulk (right hand sides of eqs.\ (\ref{eq1}) and 
(\ref{eq2})) are modelled as the corresponding energy input rates, i.e., in 
terms of $U$, $\Delta$, and the widths $\lambda_u$ and $\lambda_\theta$ of 
the kinetic and thermal boundary layers, respectively. For the thickness of 
the thermal BL we assume $\lambda_\theta = L / (2 Nu)$ and for that of the 
kinetic one $\lambda_u = L /(4 \sqrt{Re})$ as it holds in Blasius type 
layers \cite{ll87}; as for the prefactor $1/4$ cf. \cite{gro00}, Sec. 4.3. 
For very large $Ra$ the laminar BL will become turbulent  and $\lambda_u$
will show a stronger $Re$ dependence. Note that whereas the thermal 
BLs only build up at the top and bottom wall, the kinetic BL occurs
at {\it all} walls of the cell and therefore 
the contribution of $\eps_{u,BL}$
to $\eps_u$ is larger than a simple minded argument would suggest.
The two eqs.\ (\ref{eq1}) and (\ref{eq2}) then allow to calculate the two 
dependent variables $Nu$ and $Re$ as functions of the two independent ones 
$Ra$ and $Pr$.

\begin{figure}[htb]
\setlength{\unitlength}{1.0cm}
\begin{picture}(11,10)
\put(0.0,0.5)
{\epsfig{figure=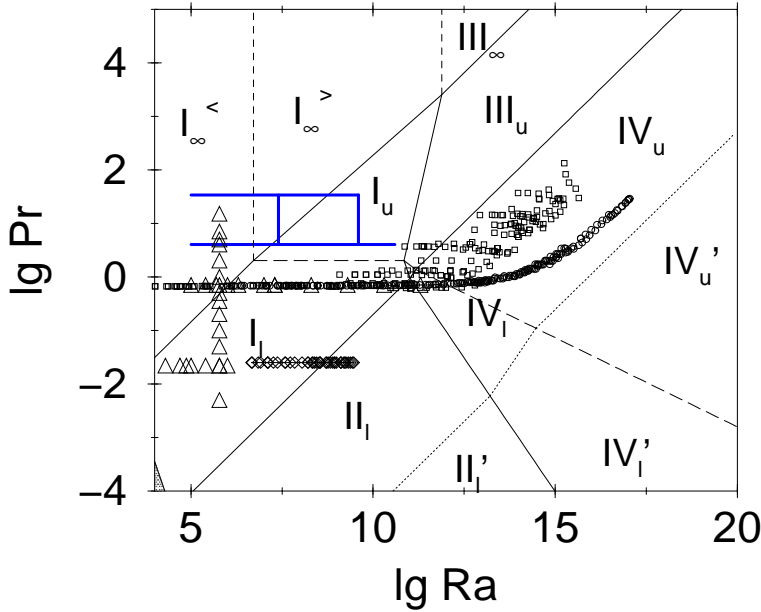,width=10cm,angle=-90}}  
\end{picture}
\caption[]{
Phase diagram in the $Ra-Pr$ plane, with data points included
where $Nu$ has
been measured or numerically calculated.
Squares show measuring points of Chavanne et al.\ \cite{cha97},
diamonds those by Cioni et al.\ \cite{cio97},
circles those by Niemela et al.\ \cite{nie00},
the very thick lines those by Xu et al.\ \cite{ahl00},
and the triangles are those points for which Verzicco and 
Camussi did full numerical simulations \cite{ver99}.
The long-dashed line is the line $\lambda_u = \lambda_\theta$. 
The thin dotted line denotes where the laminar kinetic BL becomes
turbulent. As extensively discussed in ref.\ \cite{gro00}, the exact
onset of this instability strongly depends on the prefactors used
when calculating this type of phase diagram. For this phase diagram 
the same prefactors as in ref.\ \cite{gro00} have been chosen.
}
\label{fig1}
\end{figure}

{\it Input rate modeling:}
The modeling of the 
dissipation rates on the rhs of eqs.\ (\ref{eq1}) and (\ref{eq2}) is
guided by 
the Boussinesq equations. Depending on whether the BL or the 
bulk contributions are dominant, one gets different expressions on the rhs 
of eqs.\ (\ref{eq1}), (\ref{eq2}) and
thus different relations  for $Nu$, $Re$ vs. $Ra$,
$Pr$, defining different main regimes, see ref.\ \cite{gro00} and 
fig.\ \ref{fig1}.

The rhs thermal dissipation rates depend on whether the 
kinetic BL of thickness $\lambda_u$ is within the thermal BL of thickness 
$\lambda_\theta$ ($\lambda_u < \lambda_\theta$, small $Pr$) or vice versa
($\lambda_u > \lambda_\theta$, large $Pr$). The line $\lambda_u
=\lambda_\theta$, corresponding to $Nu = 2 \sqrt{Re}$,
splits the phase diagram into a lower (small $Pr$) and an upper (large $Pr$) 
part, which we shall label by ``$\ell$" and ``$u$". 

We first consider $\lambda_u < \lambda_\theta$ (i.e., small $Pr$, regime
``$\ell$"). Then (see \ \cite{gro00}) 
\be
\eps_{u,bulk} \sim {U^3\over L} \sim {\nu^3\over L^4} Re^3,
\label{epsubulk}
\ee
\be
\eps_{u,BL} \sim \nu  {U^2 \over \lambda_u^2 } {\lambda_u \over L}
\sim {\nu^3 \over L^4 } Re^{5/2},
\label{eq4}
\ee
\be
\eps_{\theta , bulk} \sim 
{U\Delta^2 \over L}
\sim \kappa {\Delta^2 \over L^2} Pr Re,
\label{eq5n}
\ee
\be
\eps_{\theta , BL} \sim \kappa {\Delta^2 \over L^2} (Re Pr)^{1/2}.
\label{eq6}
\ee
The last expression is concluded \cite{shr90,sig94,gro00} from the heat 
transfer equation
$u_x \partial_x \theta + u_z \partial_z \theta=\kappa \partial_z^2 \theta$,
which implies ${U / L} \sim {\kappa / \lambda_\theta^2}$ giving 
$Re^{1/2} Pr^{1/2} \sim Nu $. 

If now larger $Pr$ are considered, the kinetic boundary will eventually 
exceed the thermal one, $\lambda_u > \lambda_\theta$, upper range ``$u$". The 
relevant velocity at the edge between the thermal BL and the thermal bulk now 
is less than $U$, about $U\lambda_\theta / \lambda_u$. To describe the 
transition from  $\lambda_u$ being smaller to being larger than 
$\lambda_\theta$ we introduce the function $f(x)=(1+x^n)^{-1/n}$ of the variable 
$x_\theta = \lambda_u / \lambda_\theta =  c_\theta Nu / 2\sqrt{Re} $, $f$
being 1 in the lower range ``$\ell$" (small $Pr$) and $1/x_\theta$ in ``$u$" 
(large $Pr$), respectively. $c_\theta$ is about $1$. The relevant velocity 
then is $Uf(x_\theta)$. We take $n = 4$ to characterize the sharpness of the 
transition.
  
This generalizes (5), (6) to 
\be
\eps_{\theta , bulk} 
\sim \kappa {\Delta^2 \over L^2} Pr Re f(c_\theta Nu / 2 \sqrt{Re}),
\label{eq11}
\ee
\be
\eps_{\theta , BL} 
\sim \kappa {\Delta^2 \over L^2} \sqrt{Pr Re f(c_\theta Nu / 2 \sqrt{Re})},
\label{eq11s}
\ee
while $\eps_{u,bulk}$ and $\eps_{u,BL}$ are still given by (3),(4). Introducing
(7),(8),(3),(4) into (1),(2) leads to the $Ra$,$Pr$ dependences of $Nu$,$Re$ in 
the upper regime ``$u$". The pure power laws $Nu(Ra,Pr)$ and $Re(Ra,Pr)$ in both
the {\it lower} ``$\ell$"  ($\lambda_u < \lambda_\theta$, small $Pr$) and the 
{\it upper} ``$u$" ($\lambda_u > \lambda_\theta$, large $Pr$) regimes are 
summarized in table I.

{\it Very large $Pr$ regime}:
We now extend the theory to very large Prandtl numbers. 
As long as $Ra > Ra_c =
1708$ there still is wind, even for very 
large $Pr$, as $Ra_c$ is independent of $Pr$.
However, to keep $Ra$ fixed, one has to increase
the temperature difference $\Delta$ with increasing $\nu$ or $Pr$.
$Pr \gg 1$ will
lead to a smaller and smaller  large scale wind $Re$. 
The flow will eventually
become laminar throughout the cell. $\lambda_u$ can no longer
continue to increase according to
$\lambda_u \sim  Re^{-1/2}$ with decreasing $Re$, but will
saturate to a constant value of order $L$. This is the central new point of the
very large $Pr$ regime. In \ \cite{gro00} we had assumed that this 
happens at about $Re=50$. This corresponds to a saturation at 
$\lambda_u = L/4\sqrt{Re} = L/28$. To model the smooth transition to the very 
large $Pr$ regime beyond the line $Re = 50$, i.e., if  $\lambda_u = 
L/4\sqrt{Re} $ approaches $\lambda_u = L/28$ we use the crossover function 
$g(x) = x (1+ x^n)^{-1/n}$ of the crossover variable 
$x_L = 28 \lambda_u /L = 7/\sqrt{Re}$, and again $n=4$. The function $g$ 
increases linearly, $g(x_L) = x_L$, below the transition ($x_L$ small) and is 1
in the very large $Pr$ regime with $Re \le 50$. In the above modelings for the 
local dissipations rates we have to replace each $\lambda_u$ by $g(x_L)L/28$.

The resulting formulae, given momentarily, will lead, depending on $Ra$, to 
three new regimes, valid for very large $Pr$, denoted as $I_{\infty}^{<}$, 
$I_{\infty}^{>}$, and $III_{\infty}$, see Fig.\ \ref{fig1} and  Table I.

 \begin{table}
 \begin{center}
 \begin{tabular}{|c|c|c|c|c|}
 \hline
          regime 
       &  dominance of
       &  BLs
       & $Nu$
       & $Re$
\\
\hline
         $I_l $ 
       & $\eps_{u,BL}$, $\eps_{\theta,BL}$
       & $\lambda_u < \lambda_\theta$
       & $ Ra^{1/4} Pr^{1/8} $
       & $ Ra^{1/2} Pr^{-3/4} $
\\         $I_u$ 
       & 
       & $\lambda_u > \lambda_\theta$
       & $ Ra^{1/4} Pr^{-1/12} $
       & $ Ra^{1/2} Pr^{-5/6} $
\\         $I_\infty^<$ 
       & 
       & $\lambda_u = L/4 > \lambda_\theta$
       & $ Ra^{1/3}  $
       & $  Ra^{2/3} Pr^{-1} $
\\         $I_\infty^>$ 
       & 
       & $\lambda_u = L/4 > \lambda_\theta$
       & $ Ra^{1/5}  $
       & $  Ra^{3/5} Pr^{-1} $
\\
\hline
         $II_l$
       & $\eps_{u,bulk}$, $\eps_{\theta,BL}$
       & $\lambda_u < \lambda_\theta$
       & $ Ra^{1/5} Pr^{1/5} $
       & $ Ra^{2/5} Pr^{-3/5} $
\\
\hline
         $III_u$
       &  $\eps_{u,BL}$, $\eps_{\theta,bulk}$
       &  $\lambda_u > \lambda_\theta$
       & $ Ra^{3/7} Pr^{-1/7} $
       & $ Ra^{4/7} Pr^{-6/7} $
\\
         $III_\infty$
       & 
       & $\lambda_u = L/4 > \lambda_\theta$
       & $ Ra^{1/3} $
       & $ Ra^{2/3} Pr^{-1} $
\\
\hline
         $IV_l$
       & $\eps_{u,bulk}$, $\eps_{\theta,bulk}$
       & $\lambda_u < \lambda_\theta$
       & $ Ra^{1/2} Pr^{1/2} $
       & $ Ra^{1/2} Pr^{-1/2} $
\\
         $IV_u$
       & 
       & $\lambda_u > \lambda_\theta$
       & $ Ra^{1/3}  $
       & $Ra^{4/9} Pr^{-2/3} $
\\
 \hline
 \end{tabular}
 \end{center}
\caption[]{
The power laws for $Nu$ and $Re$ of the presented theory.
The regimes $II_u$ and $III_l$ are not included as
they do not or hardly exist for the choice of prefactors.
}
\label{tab1}
 \end{table}

While eq.\ (3) 
for $\eps_{u,bulk}$ is assumed to still hold   
in the very large $Pr$ range
(where $\eps_{u,bulk}$ of course hardly contributes to $\eps_u$ 
due to the large extension of the kinetic BLs),
 eqs. (4), (5), and (6) have to be generalized. 
First generalize (4) for $\eps_{u,BL} $,
\be
\eps_{u,BL} \sim \nu {U^2 \over g(x_L) L^2} \sim {\nu^3 \over L^4 } \ \ 
{ Re^2 \over g(7/\sqrt{Re})}.
\label{eq5}
\ee
Next $\eps_{\theta,bulk}$: Above the wind velocity $U$
in (5), which sets the time
scale of the stirring, has already been generalized to  
$U f(\lambda_u / \lambda_\theta)$. This equals $U$ itself in ``$\ell$" and $U
\lambda_\theta/\lambda_u$ in the ``$u$" regimes.
Now in addition the explicit $\lambda_u$ 
is to be replaced by $g(x_L)L/28$, i.e., 
\be
\eps_{\theta,bulk} \sim \kappa{\Delta^2 \over L^2} Pr Re \ \ 
f\left( {Nu \over 14}g\left({7
\over \sqrt{Re}}\right) \right).
\label{eq10}
\ee
This simplifies for large enough $Ra$ (therefore large $f$-argument) and very 
large $Pr$ (thus large $g$-argument) to    
\be
\eps_{\theta,bulk} \sim \kappa{\Delta^2 \over L^2} \ \ {Pr Re \over Nu}.
\label{eq11nn}
\ee
Inserting (9) and (11) into the rhs of (1) and (2) leads to the new power laws
describing the heat flux and the wind velocity in the regime $III_\infty$ 
beyond $III_u$, cf. Figure 1,
\be
Nu \sim Ra^{1/3} Pr^{0}, \ \ Re \sim Ra^{2/3} Pr^{-1} .
\label{eq12}
\ee
Finally $\eps_{\theta,BL}$: In the thermal boundary layer range beyond 
$I_{\ell}$, relevant for medium $Ra$, eq.(6) stays valid, because its 
derivation  did not involve $\lambda_u$ and also $f = 1$. The
range $I_\infty^<$ \ \ has to be described by eqs.(1),(2),(9) (with $g = 1$), 
and (6), resulting in the same  power laws as in regime $III_\infty$, i.e.,
eqs.\ (\ref{eq12}). 
For $Pr$ values above  regime 
$I_u$ eq.(6) no longer holds. It originated from the heat transport
equation. There we have to use now $Uf(x_\theta)$ instead of merely $U$. 
The balance from the heat transfer equation then
reads $U f(x_\theta) / L \sim \kappa/\lambda_\theta^2$. In the 
$f$-argument $x_\theta = \lambda_u / \lambda_\theta$ 
the kinetic BL width  $\lambda_u$ and therefore the
crossover function $g$ appears, leading to  
\be
Pr Re f \left( {Nu \over 14} g({7 \over \sqrt{Re}})  \right) \sim Nu^2.
\label{eq14}
\ee
In the very large $Pr$ regime (where $g(x_L) =1$) 
above $I_u$ (where $f(x_\theta) = x_\theta^{-1}$) 
one obtains from (\ref{eq14}) the relation
$
(Re Pr)^{1/3} \sim Nu$,
valid in $I_\infty^>$. Together with (1), (2), and (9)  
one derives the
new scaling laws in the interior of $I_\infty^>$
\be
Nu \sim Ra^{1/5} Pr^{0}, \ \ Re \sim Ra^{3/5} Pr^{-1}.
\label{eq16}
\ee
The scaling behavior $Nu \sim Ra^{1/5}$ has earlier been suggested by Roberts 
\cite{rob79}. Note that in all three very large $Pr$ regimes $Nu$ does not
depend on $Pr$. Furthermore the Constantin-Doering \cite{con99} upper bound 
$Nu \le const Ra^{1/3}  (1+\log Ra)^{2/3}$, holding in the limit $Pr 
\rightarrow \infty$, is strictly fulfilled. 

\begin{figure}[htb]
\setlength{\unitlength}{1.0cm}
\begin{picture}(8,7)
\put(0.0,0.5)
{\epsfig{figure=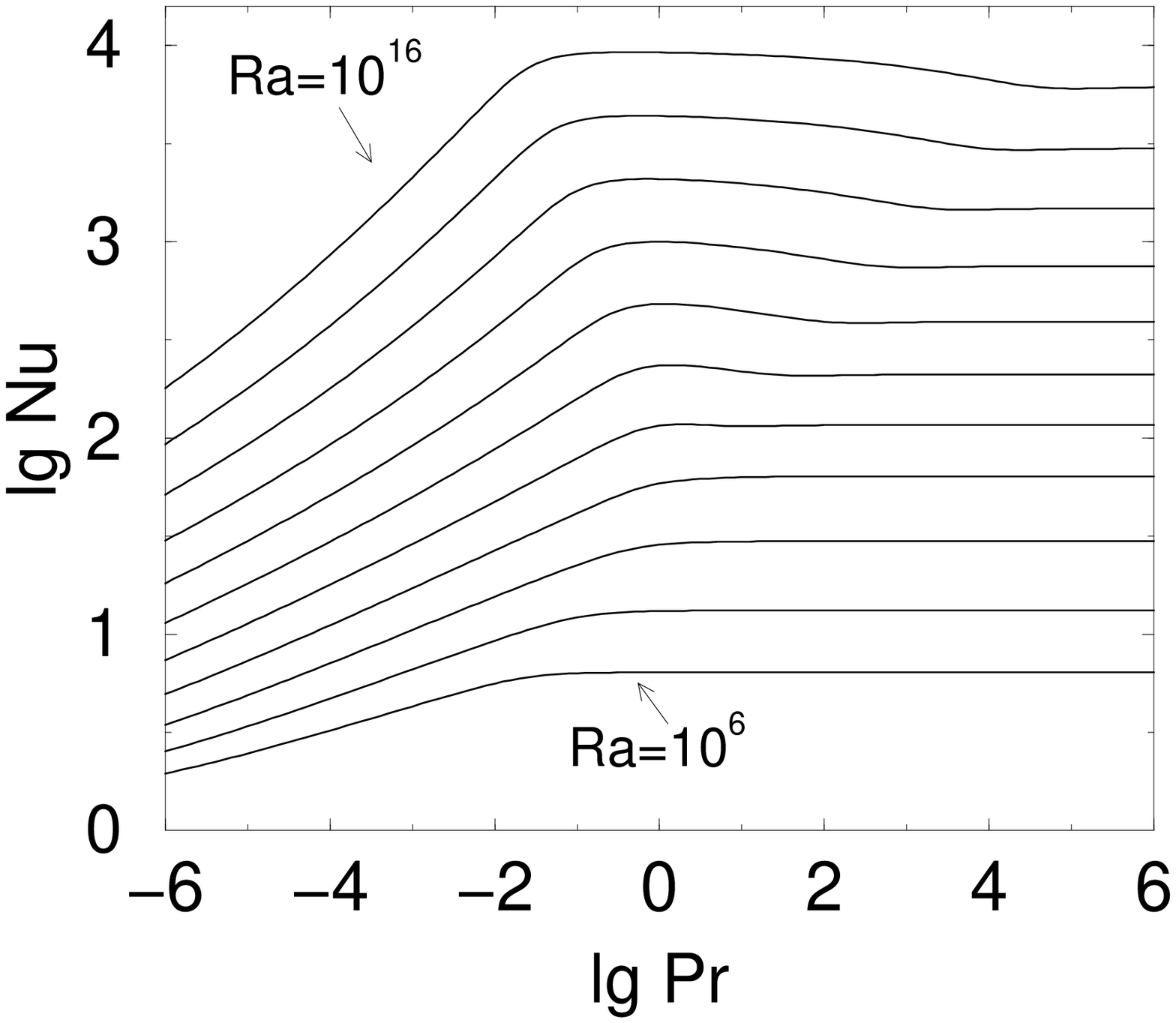,width=6.2cm,angle=-90}}
\put(8.0,0.5)
{\epsfig{figure=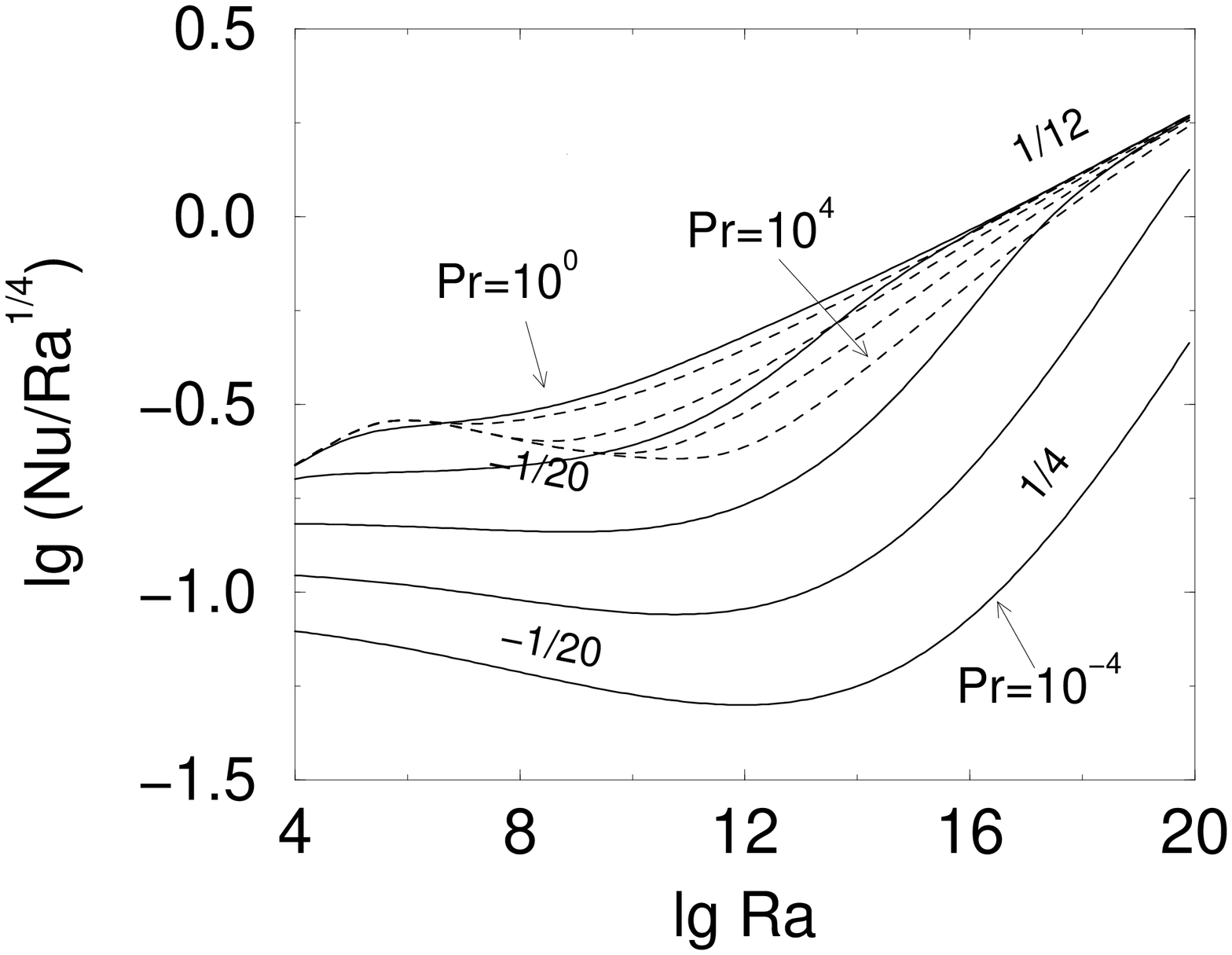,width=7cm,angle=-90}}
\end{picture}
\caption[]{
(a) $Nu$ as a function of $Pr$ according to theory for $Ra = 10^6$, 
$Ra=10^7$, \dots
to 
$Ra=10^{16}$, bottom to top.
(b) $Nu/Ra^{1/4}$ as function of $Ra$ for 
$Pr=10^{-4}$,
$Pr=10^{-3}$,
$Pr=10^{-2}$,
$Pr=10^{-1}$, and
$Pr=10^{0}$  (solid lines, bottom to top) and for 
$Pr=10^{1}$,
$Pr=10^{2}$,
$Pr=10^{3}$, and
$Pr=10^{4}$ (dashed lines, top to bottom).
}
\label{fig3}
\end{figure}

{\it $Nu$ and $Re$ in the whole parameter plane}:
Plugging now the general expressions for the local dissipation rates (3), (9), 
(10), and (\ref{eq14}) into the balance eqs.\ (\ref{eq1}) and (\ref{eq2}) finally 
results in
\be
Nu Ra Pr^{-2} = c_1 {Re^2 \over g (7/\sqrt{Re})} + c_2 Re^3,
\label{eq17}
\ee
\be
Nu = c_3 Re^{1/2} Pr^{1/2} 
\left[ 
f\left( c_\theta {Nu \over 14} g({7 \over \sqrt{Re}}) \right)
\right]^{1/2}
+ c_4 Pr Re
f\left(  c_\theta {Nu \over 14} g({7 \over \sqrt{Re}}) \right).
\label{eq18}
\ee
Here we have added dimensionless prefactors where appropriate to complete the
modeling of the dissipation rates. In ref.\ \cite{gro00} $c_1$ through
$c_5$ were already adopted to Chavanne et al.'s experimental data \cite{cha97}.
The result was
$c_1 = 1028$,
$c_2 = 9.38$,
$c_3 = 1.42$,
$c_4 = 0.0123$,
$1/c_\theta = c_5 = 2.0$, 
see eqs.\ (4.4) through (4.9) of \cite{gro00}. Note that the $c_i$ 
may depend on the aspect ratio and are not universal. 

The set of eqs.(\ref{eq17}) and (\ref{eq18}) 
is the second main result of this paper. It allows 
to calculate $Nu(Ra,Pr)$ and $Re(Ra,Pr)$ in the whole $Ra - Pr$ parameter 
space, including all 
crossovers from any regime to any neighboring one. Technically, eq. 
(\ref{eq17}) is solved for $Nu(Ra,Re,Pr)$ which then is inserted into eq.
(\ref{eq18}), 
leading to {\it one} implicit equation for $Re(Ra,Pr)$. 

All limiting, pure scaling regimes which can be derived from eqs.
(\ref{eq17}) and (\ref{eq18}) are listed in table I which extends table II 
of ref.\ \cite{gro00}, now including the three new regimes $I_\infty^<$, 
$I_\infty^>$, and $III_\infty$. The corresponding phase diagram is shown
in figure \ref{fig1}, completing that of \cite{gro00} towards  very
large $Pr$.

Though in the phase diagram we have drawn lines to indicate transitions between
the regimes, defined by either
$\eps_{u,BL} = \eps_{u,bulk}$ or
$\eps_{\theta,BL} = \eps_{\theta,bulk}$ or
$\lambda_u =\lambda_\theta$ or $\lambda_u = L/28$, 
we note that the crossovers are nothing at all but sharp. All transitions are 
smeared out over broad ranges, the more, the more similar the 
scaling exponents of the neighboring regimes are.

{\it Discussion  of $Nu(Pr)$ and $Nu(Ra)$:}
 The functions $Nu(Pr)$ (for fixed values of $Ra$) and 
$Nu(Ra)/Ra^{1/4}$ (for fixed values of $Pr$) resulting from eqs.\
(\ref{eq17}) - (\ref{eq18}) are shown 
in figure \ref{fig3}.
Indeed, for $Ra\approx 10^6 - 10^9$ 
the Nusselt number saturates with increasing
$Pr$ and the regime $I_u$ with $Nu\sim Pr^{-1/12}$ is suppressed, in full
agreement with the experimental and numerical results 
\cite{ahlers_sb00,ver99,her00}.
However,
for larger $Ra$ beyond $\approx 10^{11}$ 
the theory predicts a maximum for the curve
$Nu(Pr)$. The decreasing branches of the curve for $Ra\approx 10^{12}$
and $Pr\approx 50$ may be consistent with above mentioned experimental
results by 
Ashkenazi and Steinberg \cite{ash99}. This observation may resolve the
apparent descrepancy between the Ahlers et al.\ and the Steinberg et al.\
data: Simply different regimes in the $Ra-Pr$ phase space are probed.

The importance of transition ranges is highlighted in fig.\ \ref{fig3}b.
E.g., in regime $II_l$ with $Nu\sim Ra^{1/5}$ (for fixed $Pr$) 
the corresponding scaling exponent $1/5-1/4 = -1/20$ only becomes
observable for very small $Pr\approx 10^{-3} - 10^{-4}$. At
$Pr=10^{-2}$ the roughly four decades of regime $II_l$ (see figure 1)
between regime $I_l$ and $IV_l$, which both have a stronger $Ra$ 
dependence of $Nu$, are not sufficient to reveal the scaling exponent.
And at $Pr=10^{-1}$ the roughly 2.5 decades of regime $II_l$ are 
{\it nowhere} sufficient to lead to a local scaling exponent
$dNu/dRa$ smaller than $1/4$!

Similarly, for $Pr=1$ only for very large $Ra \ageq 10^{15}$ pure
scaling $Nu\sim Ra^{1/3}$ is revealed. For smaller $Ra$ regimes $I_u$ 
and $I_l$ with $Nu\sim Ra^{1/4}$ strongly
contribute, resulting in an effective local scaling exponent 
(increasing with $Ra$) in the
range between $0.28$ (``$2/7$'') and $0.31$,
just as observed in experiment \cite{cas89,sig94,cha97,nie00}.

\begin{figure}[htb]
\setlength{\unitlength}{1.0cm}
\begin{picture}(8,7)
\put(0.0,0.5)
{\epsfig{figure=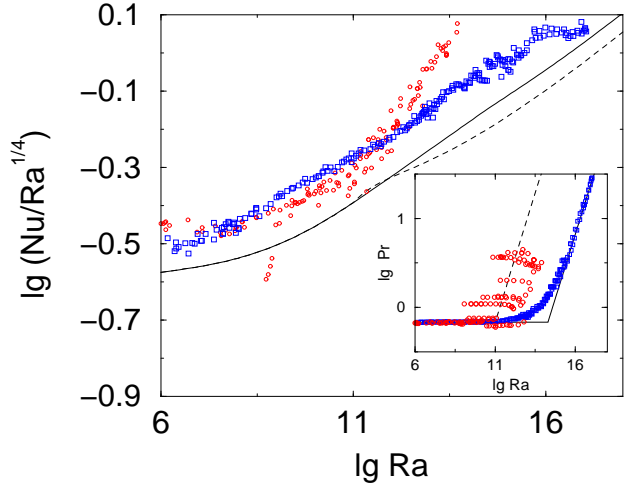,width=8.0cm,angle=-90}}
\end{picture}
\caption[]{
$Nu(Ra,Pr(Ra))/Ra^{1/4}$ 
along the curve $Pr(Ra)$ given by the experimental restrictions of 
either the Niemela et al.\ experiment \cite{nie00} (boxes in the inset
and solid line in the main figure; the experimental data itself are
shown as boxes)
or  the Chavanne et al.\ experiment \cite{cha97} (circles in the inset
and dashed line in the main figure; the experimental data itself are
shown as circles). The simple parameterizations on which the curves in the
main figures are based are also indicated. 
}
\label{fig4}
\end{figure}

{\it Direct comparison to experiment:} 
Finally, let us point out that the present approach simplifies the 
comparison with experimental data: In experiment, it is hard to vary either
$Ra$ or $Pr$ over many decades {\it and} at the same time to keep the other
variable fixed. So most measurements are done along {\it curved} lines in
the phase space figure \ref{fig1}, mixing the $Ra$ and $Pr$ dependences.
Now eqs.\ (\ref{eq17}) and (\ref{eq18}) allow to calculate $Nu$ and $Re$ 
along such a curve $Pr(Ra)$ given by the experimental restrictions, which
connect $Ra$ and $Pr$. E.g., figure \ref{fig4} shows $Nu(Ra, Pr(Ra))$ with
$Pr(Ra)$ as resulting from the Niemela et al.\ \cite{nie00} 
and the Chavanne et al.\ \cite{cha97} 
experiments.

The present theory suggests that beyond $Ra\approx 10^{12}$ the Nusselt
number in the Chavanne et al.\ experiment \cite{cha97} should be 
{\it smaller} than in the Niemela et al.\ experiment \cite{nie00},
in contrast to what is found. This result is also in contrast to
our former speculation of ref.\ \cite{gro00}, that the strong
$Nu \sim Ra^{3/7}$ dependence of regime $III_u$, to which the Chavanne
et al.\ data are closer, would account for their experimental findings.
What we had overseen is that the larger $Pr$ numbers of the Chavanne et
al.\ experiment (for fixed $Ra$) and the $Nu\sim Pr^{-1/7}$ dependence
of regime $III_u$ overcompensates the strong $Ra$ dependence of
$Nu$ for the Chavanne et al.\ data close to regime $III_u$. 

This present theory cannot resolve the paradox between the Chavanne
et al.\ and the Niemela et al.\ data.
Possibly, different temperature boundary conditions have been applied.
Possibly, Chavanne et al.\ have already observed the transition from
a laminar to a turbulent kinetic BL (dotted line in the phase diagram
fig.\ \ref{fig1}).

{\bf Acknowledgement:} 
This research work was prompted by the experiments of
G.\ Ahlers \cite{ahlers_sb00}. We thank him for sharing
his results with us prior to publication and  
for various discussions.


\begin{thebibliography}{10}

\bibitem{gro00}
S. Grossmann and D. Lohse, J. Fluid. Mech. {\bf 407},  27  (2000).

\bibitem{cas89}
B. Castaing {\it et~al.}, J. Fluid Mech. {\bf 204},  1  (1989).

\bibitem{sig94}
E.~D. Siggia, Annu. Rev. Fluid Mech. {\bf 26},  137  (1994).

\bibitem{cio97}
S. Cioni, S. Ciliberto, and J. Sommeria, J. Fluid Mech. {\bf 335},  111
  (1997).

\bibitem{cha97}
X. Chavanne {\it et~al.}, Phys. Rev. Lett. {\bf 79},  3648  (1997).

\bibitem{nie00}
J. Niemela, L. Skrebek, K.~R. Sreenivasan, and R. Donelly, Nature {\bf 404},
  837  (2000).

\bibitem{ash99}
S. Ashkenazi and V. Steinberg, Phys. Rev. Lett. {\bf 83},  3641  (1999).


\bibitem{ahl00}
X. Xu, K.~M.~S. Bajaj, and G. Ahlers, Phys. Rev. Lett. {\bf 84},  4357  (2000).

\bibitem{ahlers_sb00}
G. Ahlers, 2000, seminar at the ITP-Workshop in Santa Barabara.

\bibitem{ver99}
R. Verzicco and R. Camussi, J. Fluid Mech. {\bf 383},  55  (1999).

\bibitem{her00}
J. Herring and R. Kerr, Preprint (2000).

\bibitem{ll87}
L.~D. Landau and E.~M. Lifshitz, {\em Fluid Mechanics} (Pergamon Press, Oxford,
  1987).

\bibitem{shr90}
B.~I. Shraiman and E.~D. Siggia, Phys. Rev. A {\bf 42},  3650  (1990).

\bibitem{rob79}
G.~O. Roberts, Geophys. Astrophys. Fluid Dyn. {\bf 12},  235  (1979).

\bibitem{con99}
P. Constantin and C. Doering, J. Stat. Phys. {\bf 94},  159  (1999).

\end{thebibliography}

\end{document}